\documentclass[%
reprint,
amsmath,amssymb,
aps,
prb,
]{revtex4-2}
\usepackage{graphicx}
\usepackage{CJKutf8}
\usepackage{epstopdf}
\usepackage{subfigure}
\usepackage{pifont}
\usepackage{bm}
\usepackage{xcolor}
\usepackage[colorlinks=true,linkcolor=blue,anchorcolor=blue,citecolor=blue,urlcolor=blue]{hyperref}
\begin{document}
\begin{CJK}{UTF8}{gbsn}
\preprint{APS/123-QED}
\title{Layer-Resolved Impurity States Reveal Competing Pairing Mechanisms in Trilayer Nickelate Superconductor La$_4$Ni$_3$O$_{10}$
}
\author{Suyin Zheng\textsuperscript{1} and Tao Zhou\textsuperscript{1,2}}
\email{tzhou@scnu.edu.cn}
\affiliation{ \textsuperscript{1}Guangdong Basic Research Center of Excellence for Structure and Fundamental Interactions of Matter, \\
Guangdong Provincial Key Laboratory of Quantum Engineering and Quantum Materials, \\
School of Physics, South China Normal University, Guangzhou 510006, China \\
\textsuperscript{2}Guangdong-Hong Kong Joint Laboratory of Quantum Matter, \\
Frontier Research Institute for Physics, South China Normal University, Guangzhou 510006, China}

\begin{abstract}
Trilayer Ruddlesden-Popper nickelate superconductor $\mathrm{La}_4 \mathrm{Ni}_3 \mathrm{O}_{10}$ has generated considerable interest due to its unconventional superconductivity and complex electronic structure. Notably, $\mathrm{La}_4 \mathrm{Ni}_3 \mathrm{O}_{10}$ features a mixed Ni valence state and an asymmetric trilayer configuration, leading to distinct quasiparticle distributions and local density of states (LDOS) between the inner and outer NiO$_2$ planes. In this work, we investigate impurity-induced states in $\mathrm{La}_4 \mathrm{Ni}_3 \mathrm{O}_{10}$ using a two-orbital model combined with $T$-matrix formalism, focusing on the contrasting roles of intra- and interlayer pairing channels. Our self-consistent mean-field analysis reveals that interlayer pairing results in partially gapless Fermi surfaces, with unpaired quasiparticles concentrated in the outer layers and a pronounced low-energy LDOS. We demonstrate that impurity effects vary significantly depending on both the pairing symmetry and impurity location: interlayer-dominant pairing produces sharp resonance states when impurities are in the inner layer, whereas impurities in the outer layer lead to in-gap enhancements without sharp resonances; in contrast, intralayer-dominant pairing generally yields increased in-gap LDOS without sharp impurity resonances, regardless of impurity position. These findings suggest that single-impurity spectroscopy can serve as a powerful probe to distinguish between competing superconducting pairing mechanisms in trilayer nickelates and highlight the rich physics arising from their multilayer structure.

\end{abstract}

\maketitle

\section{Introduction}

Recent breakthroughs in Ruddlesden-Popper (RP) phase bilayer nickelate superconductors, such as $\mathrm{La}_3\mathrm{Ni}_2\mathrm{O}_7$ with a critical temperature ($T_c$) reaching up to $80~\mathrm{K}$ under applied pressure, have opened new directions for investigating the mechanisms of high-temperature superconductivity~\cite{ref1,ref2}. Inspired by trilayer cuprates, which host the highest $T_c$ values among all cuprate families~\cite{ref3,ref4}, researchers have subsequently focused on trilayer nickelates, motivated by their promising superconducting properties~\cite{ref5}.

Superconductivity was soon confirmed in the trilayer RP phase nickelate $\mathrm{La}_4\mathrm{Ni}_3\mathrm{O}_{10}$, with $T_c$ in the range of $20-30$ K~\cite{ref6,ref7,ref8,ref9}. This compound exhibits a mixed valence state of $\mathrm{Ni}^{2.67+}$ and undergoes a pressure-induced structural transition to a tetragonal phase, analogous to that observed in $\mathrm{La}_3\mathrm{Ni}_2\mathrm{O}_7$~\cite{ref9}.

Both $\mathrm{La}_3\mathrm{Ni}_2\mathrm{O}_7$ and $\mathrm{La}_4\mathrm{Ni}_3\mathrm{O}_{10}$ can be effectively described by a two-orbital model incorporating the $d_{x^2-y^2}$ and $d_{z^2}$ orbitals~\cite{PhysRevLett.131.126001,ref12}. The normal state band structure naturally leads to a competition or coexistence between intralayer and interlayer pairing: strong intralayer hopping via the $d_{x^2-y^2}$ orbital and strong interlayer hopping via the $d_{z^2}$ orbital suggest that both intralayer and interlayer couplings are possible, potentially resulting in competing superconducting channels~\cite{ref20,ref21,ref22,ref24,ref25,Luo2024,ref26,ref27,ref28,PhysRevB.109.165154,PhysRevB.109.L081105,ref29,ref30,ref31,ref32,ref33,ref34,ref35,PhysRevB.111.094504,PhysRevB.111.104505,rm9g-8lm1,250607741}. To date, the superconducting pairing symmetry of $\mathrm{La}_4\mathrm{Ni}_3\mathrm{O}_{10}$ remains an important and unresolved issue~\cite{ref11,ref13,ref14,ref15,ref16,ref17,ref18,ref19,ref10,Qiong2024}.

Distinct from the symmetric bilayer structure of $\mathrm{La}_3\mathrm{Ni}_2\mathrm{O}_7$, the trilayer configuration of $\mathrm{La}_4\mathrm{Ni}_3\mathrm{O}_{10}$ is inherently asymmetric~\cite{ref10,Qiong2024}. The quasiparticle distributions near the Fermi level differ between the inner and outer $\mathrm{NiO}_2$ planes. In the superconducting state, interlayer pairing leaves certain quasiparticles near the Fermi level unpaired. These unpaired quasiparticles predominantly reside in the outer layers, resulting in partially gapless Fermi surfaces. Consequently, the local density of states (LDOS) for the outer and inner layers differ significantly, with a larger low-energy LDOS present in the outer layers~\cite{ref10}.

Impurity effects have long served as powerful probes of pairing symmetry in unconventional superconductors, including cuprates~\cite{ref36,ref37}, iron-based superconductors~\cite{ref38,ref39,ref40,ref41,ref42}, and other systems~\cite{ref43,ref44,ref45,ref46,ref47,ref48,ref49,ref50,ref51,ref52,PhysRevB.111.174525}. These studies have provided both theoretical and experimental insights into impurity-induced states, combining calculations of the LDOS with scanning tunneling microscopy (STM) measurements~\cite{ref36,ref37,ref38,ref39,ref40,ref41,ref42}. 
Theoretically, impurity effects in unconventional superconductors can be systematically studied using the $T$-matrix formalism, where impurity-induced bound states are closely related to the behavior of the $T$-matrix denominator. When both the real and imaginary parts of the denominator approach zero, the $T$-matrix amplitude is enhanced, resulting in low-energy resonance states~\cite{ref36}. 

Generally, at low energies, the imaginary part of the $T$-matrix denominator is determined by the bare LDOS in the absence of impurities. In conventional superconductors, this bare LDOS is typically suppressed due to the presence of a superconducting gap, so impurity-induced resonance states are mainly governed by the real part of the denominator, which is closely tied to the sign structure of the pairing order parameter~\cite{PhysRevB.111.174525}.

However, for $\mathrm{La}_4\mathrm{Ni}_3\mathrm{O}_{10}$, the low-energy bare LDOS is not necessarily small. Specifically, for interlayer pairing, a substantial number of unpaired quasiparticles are localized in the outer layers, resulting in a relatively large low-energy LDOS for these layers, while the inner layer quasiparticles are almost gapped~\cite{ref10}. Therefore, in the interlayer-dominant pairing scenario, impurity-induced resonance states may strongly depend on the impurity position. In contrast, for intralayer-dominant pairing, the bound states are expected to be less sensitive to the impurity position. This distinction suggests that single-impurity effects could serve as a useful probe to discriminate between interlayer and intralayer pairing mechanisms.

Motivated by these considerations, in this paper we investigate the single-impurity effects in $\mathrm{La}_4\mathrm{Ni}_3\mathrm{O}_{10}$ using a two-orbital model and the $T$-matrix method. Our results reveal that, due to the trilayer structure of $\mathrm{La}_4\mathrm{Ni}_3\mathrm{O}_{10}$, single-impurity effects differ markedly between intralayer-dominant and interlayer-dominant pairing states.

The remainder of this paper is organized as follows: Section~II presents the model and the $T$-matrix formalism; Section~III discusses the numerical results; and Section~IV provides a summary.


\section{MODEL AND FORMALISM}
We investigate the trilayer nickelate system using a two-orbital model defined on a trilayer square lattice. The Hamiltonian consists of both tight-binding and interaction terms:
\begin{equation}
H = -\sum_{l, l^{\prime}} \sum_{{\bf i j} \tau \tau^{\prime} \sigma} t_{{\bf i j} \tau \tau^{\prime}}^{l, l^{\prime}} c_{{\bf i} \tau \sigma}^{l \dagger} c_{{\bf j} \tau^{\prime} \sigma}^{l^{\prime}} + H_{I},
\end{equation}
where $l/l' =1,2,$ or $3$, denoting layer indices, $\tau/\tau'$ labels orbital index with $\tau/\tau'=x$ or $z$ representing the $d_{x^2 - y^2}$ or $d_{z^2}$ orbital, respectively.
 \( \sigma \) is the spin index. ${\bf i}$ and ${\bf j}$ are lattice sites. 
 
 The pairing interaction term \( H_{I} \) describes an effective exchange interaction and is given by
\begin{equation}
	H_I =-\sum_{l,l'} \sum_{{\bf ij}\tau}\sum_{\sigma,\sigma^\prime} V_{{\bf ij}\tau}^{l,l'}  c_{{\bf i}\tau\sigma}^{l\dagger} c_{{\bf j}\tau\sigma^\prime}^{l'\dagger}c_{{\bf i}\tau\sigma^\prime}^{l'} c_{{\bf j}\tau\sigma}^{l} .
\end{equation}

Superconducting order parameters are introduced at the mean-field level. The intralayer and interlayer pairings are defined, respectively, as
\(\Delta_{{\bf i j} \tau}^{l, l} = \frac{V_{{\bf ij} \tau}^{l, l}}{2} \langle c_{{\bf i} \tau \uparrow}^{l} c_{{\bf j} \tau \downarrow}^{l} - c_{{\bf i} \tau \downarrow}^{l} c_{{\bf j} \tau \uparrow}^{l} \rangle\) and \(\Delta_{{\bf ii} \tau}^{l, l^{\prime}} = \frac{V_{{\bf ii} \tau}^{l, l^{\prime}}}{2} \langle c_{{\bf i} \tau \uparrow}^{l} c_{{\bf i} \tau \downarrow}^{l^{\prime}} - c_{{\bf i} \tau \downarrow}^{l} c_{{\bf i} \tau \uparrow}^{l^{\prime}} \rangle\).

Applying Fourier transformation, the Hamiltonian takes the form
\begin{equation}
H = \sum_{\mathbf{k}} \hat{\Psi}_{\mathbf{k}}^{\dagger} \hat{H}_{\mathbf{k}} \hat{\Psi}_{\mathbf{k}},
\end{equation}
with the $12$-component Nambu spinor:
\begin{align}
\begin{array}{l}
\hat{\Psi}_{\mathbf{k}}=(c_{\mathbf{k} x \uparrow}^{1}, c_{\mathbf{k} x \uparrow}^{2}, c_{\mathbf{k} x \uparrow}^{3}, c_{\mathbf{k} z \uparrow}^{1}, c_{\mathbf{k} z \uparrow}^{2}, c_{\mathbf{k} z \uparrow}^{3}, \\c_{-\mathbf{k} x \downarrow}^{1 \dagger}, c_{-\mathbf{k} x \downarrow}^{2 \dagger}, c_{-\mathbf{k} x \downarrow}^{3 \dagger}, c_{-\mathbf{k} z \downarrow}^{1 \dagger}, c_{-\mathbf{k} z \downarrow}^{2 \dagger}, c_{-\mathbf{k} z \downarrow}^{3 \dagger} )^{T}.\end{array}
\end{align}
The matrix
 \( \hat{H}_{\mathbf{k}} \) is of dimension \( 12 \times 12 \) and has the block structure:
\begin{equation}
\hat{H}_{\mathbf{k}} = \begin{bmatrix}
\hat{H}_{0}(\mathbf{k}) & \hat{H}_{\Delta}(\mathbf{k}) \\
\hat{H}_{\Delta}^{\dagger}(\mathbf{k}) & -\hat{H}_{0}(\mathbf{k})
\end{bmatrix},
\end{equation}
where \( \hat{H}_0(\mathbf{k}) \) represents the normal state part and \( \hat{H}_{\Delta}(\mathbf{k}) \) represents the pairing matrix. The explicit form of \( \hat{H}_{\Delta}(\mathbf{k}) \) is:
\begin{equation}
\hat{H}_{\Delta}(\mathbf{k}) = 
\begin{pmatrix}
\widetilde{\Delta}_{x}^{11} & \Delta_{x}^{12} & 0 & 0 & 0 & 0 \\
\Delta_{x}^{12} & \widetilde{\Delta}_{x}^{22} & \Delta_{x}^{23} & 0 & 0 & 0 \\
0 & \Delta_{x}^{23} & \widetilde{\Delta}_{x}^{33} & 0 & 0 & 0 \\
0 & 0 & 0 & \widetilde{\Delta}_{z}^{11} & \Delta_{z}^{12} & 0 \\
0 & 0 & 0 & \Delta_{z}^{12} &\widetilde{\Delta}_{z}^{22} & \Delta_{z}^{23} \\
0 & 0 & 0 & 0 & \Delta_{z}^{23} &\widetilde{\Delta}_{z}^{33}
\end{pmatrix}.
\end{equation}
Here, the diagonal elements correspond to intralayer pairing, while the off-diagonal elements represent interlayer pairing components.

For intralayer pairing, we consider the nearest-neighbor interaction characterized by \( V_{\tau \parallel} = V^{l,l}_{{\bf ij}\tau} \). 
 Self-consistent calculations indicate that the system favors extended 
$s$-wave pairing symmetry with $\Delta_{{\bf i j} \tau}^{l, l}\equiv \Delta_{\tau}^{ll}$ for nearest-neighbor sites ${\bf i}$ and ${\bf j}$. 
In momentum space, the intralayer pairing function is \( \widetilde{\Delta}_{\tau}^{ll} = 2\Delta_{\tau}^{ll} (\cos k_x + \cos k_y) \). The pairing order parameter $\Delta_{\tau}^{ll}$ is given by,
   
 \begin{equation}
 	\Delta_{\tau}^{ll} = \frac{V_{\tau \|}}{2 N} \sum_{\mathbf{k}} \left( \cos k_{x} + \cos k_{y} \right) \langle c^l_{{\bf k}\tau\uparrow} c^l_{{-\bf k}\tau\downarrow} \rangle,
 \end{equation}
 where $N$ is the number of the lattice sites per layer. 

By diagonalizing the Hamiltonian matrix $ \hat{H}_{\mathbf{k}} $, Eq.(7) can be rewritten as,
\begin{equation}
\Delta_{\tau}^{ll} = \frac{V_{\tau \|}}{4 N} \sum_{n \mathbf{k}} \left( \cos k_{x} + \cos k_{y} \right) u^{*}_{m n} (\mathbf{k}) u_{m+6, n} (\mathbf{k}) \tanh \frac{\beta E_{n} (\mathbf{k})}{2}.
\end{equation}
where $E_{n} (\mathbf{k})$ are the eigenvalues of the Hamiltonian and $u_{m n} (\mathbf{k})$ are the corresponding eigenvector components, with $m=l$ or $l+3$ for the $d_{x^2 - y^2}$ or $d_{z^2}$ orbital, respectively.

For the interlayer pairing, the interaction parameter is $ V_{\tau \perp}=V_{{\bf ii}\tau}^{l, l+1} $ ($l=1$ or $2$). The interlayer pairing order parameter $ \Delta_{\tau \perp} $ 
is expressed as $\Delta^{l,l+1}_{\tau}=\frac{V_{\tau \perp}}{N} \sum_{\mathbf{k}}  \langle c^l_{{\bf k}\tau\uparrow} c^{l+1}_{{-\bf k}\tau\downarrow} \rangle$.
After diagonalization, this becomes:
\begin{equation}
\Delta^{l,l+1}_{\tau} = \frac{V_{\tau \perp}}{2N} \sum_{n, \mathbf{k}} u^{*}_{m n} (\mathbf{k}) u_{m+7, n} (\mathbf{k}) \tanh \frac{\beta E_{n} (\mathbf{k})}{2} .
\end{equation}

In our trilayer system, the first and third layers  ($l=1$ and $l=3$ are structurally equivalent outer layers, while the second is the inner layer. We define the intralayer order parameters as $\Delta_{\tau}^{O}=\Delta_{\tau}^{11}=\Delta_{\tau}^{33}$ for the outer layers and
$\Delta_{\tau}^{I}=\Delta_{\tau}^{22}$ for the inner layer. The interlayer order parameter for orbital
 $\tau$ is $\Delta_{\tau\perp}=\Delta_{\tau}^{12}=\Delta_{\tau}^{23}$.

To study impurity effects, we adopt the $ T $-matrix approach. The bare Green's function without the impurity [$\hat{G}_0(\mathbf{k}, \omega)$] is a $12\times 12$ matrix, calculated through diagonalizing the Hamiltonian matrix, with the elements being expressed as 
\begin{equation}
G_{ij}^{0}(\mathbf{k}, \omega) = \sum_{n} \frac{u_{in}(\mathbf{k}) u_{jn}^{*}(\mathbf{k})}{\omega - E_{n}(\mathbf{k}) + i\Gamma}.
\end{equation}

For a single nonmagnetic impurity located at site ${\bf r_0}= (0,0) $ in layer $l$, the impurity contribution to the Hamiltonian is given by
\begin{equation}
H_{imp}=V_i \sum_{\tau\sigma} c^{l\dagger}_{{\bf r_0}\tau\sigma}c^{l}_{{\bf r_0}\tau\sigma},
\end{equation}
where $V_i$ is the impurity strength.

Eq.(11) can be written in matrix form as 
 $H_{imp}=\hat{\Psi}^\dagger_{\bf r_0}\hat{U} \hat{\Psi}_{\bf r_0}$, with
\begin{align}
\begin{array}{l}
\hat{\Psi}_{\mathbf{r_0}}=(c_{\mathbf{r_0} x \uparrow}^{1}, c_{\mathbf{r_0} x \uparrow}^{2}, c_{\mathbf{r_0} x \uparrow}^{3}, c_{\mathbf{r_0} z \uparrow}^{1}, c_{\mathbf{r_0} z \uparrow}^{2}, c_{\mathbf{r_0} z \uparrow}^{3}, \\c_{\mathbf{r_0} x \downarrow}^{1 \dagger}, c_{\mathbf{r_0} x \downarrow}^{2 \dagger}, c_{\mathbf{r_0} x \downarrow}^{3 \dagger}, c_{\mathbf{r_0} z \downarrow}^{1 \dagger}, c_{\mathbf{r_0} z \downarrow}^{2 \dagger}, c_{\mathbf{r_0} z \downarrow}^{3 \dagger} )^{T}.\end{array}
\end{align} 
Here, $\hat{U}$ is a $12\times 12$ diagonal impurity potential matrix.
Its only nonzero elements are given by $ U_{ll} = U_{l+3,l+3} = V_i $ and $ U_{l+6,l+6} = U_{l+9,l+9} = -V_i $. 

The $ T $-matrix is defined as~\cite{ref36}
\begin{equation}
\hat{T}(\omega) = \left[ \hat{I} - \hat{U} \hat{G}_0(\omega) \right]^{-1} \hat{U},
\end{equation}
where $\hat{G}_0(\omega)=\frac{1}{N} \sum_{\bf k}\hat{G}_0(\mathbf{k}, \omega) $, and $ \hat{I} $ is the $ 12 \times 12 $ identity matrix.

The full Green's function in real space is then given by
\begin{equation}
\hat{G}(\mathbf{r}, \mathbf{r}', \omega) = \hat{G}_0(\mathbf{r}, \mathbf{r}', \omega) + \hat{G}_0(\mathbf{r}, {\bf r_0}, \omega) \hat{T}(\omega) \hat{G}_0({\bf r_0}, \mathbf{r}', \omega),
\end{equation}
where $ \hat{G}_0(\mathbf{r}, \mathbf{r}', \omega) $ is obtained via Fourier transformation:
\begin{equation}
\hat{G}_0(\mathbf{r}, \mathbf{r}', \omega) = \frac{1}{N} \sum_{\mathbf{k}} \hat{G}_0(\mathbf{k}, \omega) e^{i \mathbf{k} \cdot (\mathbf{r} - \mathbf{r}')}.
\end{equation}

The LDOS at site $ \mathbf{r} $ in layer $ l $ is then calculated by
\begin{equation}
\begin{aligned}
\rho^{l}(\mathbf{r}, \omega) = & -\frac{1}{\pi} \operatorname{Im} \sum_{p=0}^{1} \left[ G_{l+3p,\, l+3p}(\mathbf{r}, \mathbf{r}, \omega) \right. \\
& \left. + G_{l+3p+6,\, l+3p+6}(\mathbf{r}, \mathbf{r}, -\omega) \right].
\end{aligned}
\end{equation}

Our calculations use eV as the energy unit. 
The hopping parameters \( t_{{\bf i j} \tau \tau^{\prime}}^{l, l^{\prime}} \) are adopted from Ref.~\cite{ref12}.
We set the impurity potential to 
 $V_i=10$~\cite{supp}, and to examine the LDOS near the impurity, we choose ${\bf r}=(1,0)$. Other parameters are set as $\beta=10^{5}$ and $\Gamma=0.004$.

\section{RESULTS AND DISCUSSION}
\begin{figure}
    \centering
    \includegraphics[width=0.9\linewidth]{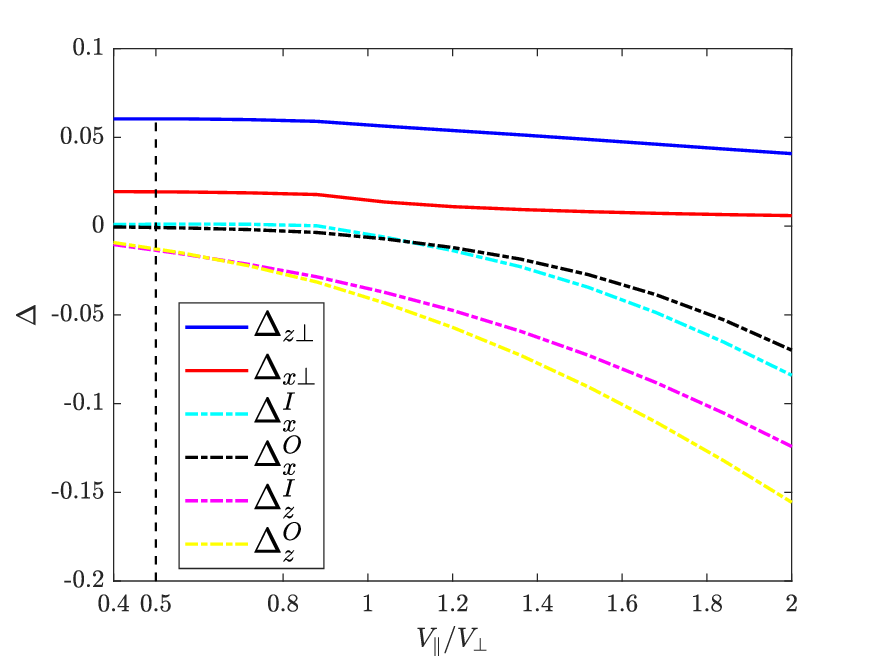}
    \caption{Self-consistent mean-field results for the superconducting order parameters as a function of the intralayer pairing potential, with the interlayer pairing potential fixed at $0.8$.}
    \label{fig1}
\end{figure}

We begin by examining the nature of the superconducting pairing within our system. Typically, the pairing interactions are closely linked to the underlying hopping parameters. In our model, the most significant hopping term is the interlayer hopping for the $d_{z^2}$ orbital, characterized by $|t_{\perp}^z| = 0.738$, followed by the intralayer nearest-neighbor hopping amplitudes for the $d_{x^2-y^2}$ orbital in the outer and inner layers, with values approximately $0.511$ and $0.521$, respectively~\cite{ref12}. Several studies have emphasized the crucial influence of Hund's coupling between the $d_{x^2 - y^2}$ and $d_{z^2}$ orbitals, noting its role in boosting the superconducting pairing strength~\cite{ref27,ref28,PhysRevB.109.165154,PhysRevB.109.L081105}. Due to the sizable Hund's interaction, these orbitals can effectively share pairing correlations, enhancing the overall pairing tendency~\cite{ref27,ref28,PhysRevB.109.165154,PhysRevB.109.L081105}. Consequently, we assume equivalent pairing interactions for both orbitals, setting $V_{\parallel}^x = V_{\parallel}^z = V_{\parallel}$ and $V_{\perp}^x = V_{\perp}^z = V_{\perp}$ in our subsequent analysis.

Fig.~\ref{fig1} displays the self-consistent calculations of the superconducting order parameters as functions of the intralayer pairing potential, with the interlayer pairing fixed at $0.8$. Within the $t-J$ model framework, the pairing interaction strength $V$ correlates with the superexchange interaction $J$, which scales as $J \approx 4t^2/U$, implying that $V$ is proportional to the square of the hopping amplitude, i.e., $V \propto t^2$. Based on this relation, the ratio of the intralayer to interlayer pairing potentials is estimated as $V_\parallel / V_\perp \approx 0.5$. As shown in the figure, under this ratio, the interlayer pairing component dominates and is notably larger than the intralayer order parameters. Increasing $V_\parallel$ results in a slight reduction in the interlayer order parameter while enhancing the intralayer one. When the ratio $V_\parallel / V_\perp$ exceeds approximately $1.2$, the intralayer pairing surpasses the interlayer component. It is important to note that the signs of the interlayer and intralayer order parameters remain opposite throughout these regimes.

In the subsequent analysis, we explore the effects of a single impurity across various parameter configurations. For cases where interlayer pairing is dominant, we consider both a purely interlayer pairing state $(V_\parallel, V_\perp) = (0, 0.8)$ and a mixed state with coexistence $(V_\parallel, V_\perp) = (0.6, 0.8)$. Conversely, for predominantly intralayer pairing scenarios, we examine the pure intralayer pairing $(V_\parallel, V_\perp) = (0.8, 0)$, as well as a coexistence regime $(V_\parallel, V_\perp) = (0.8, 0.4)$.

\begin{figure}
	\centering
	\includegraphics[width=\linewidth]{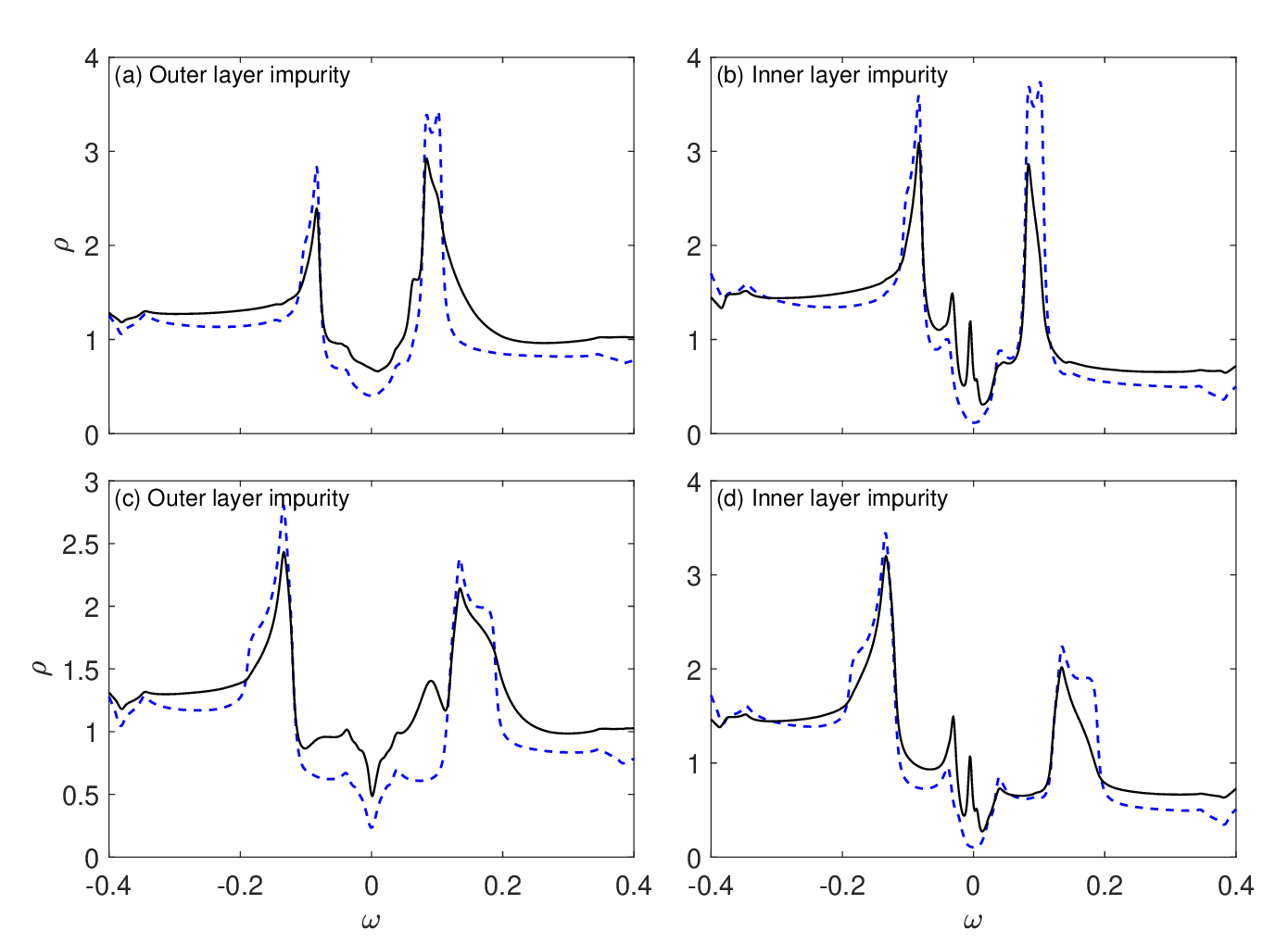}
	\caption{LDOS spectra in the case of dominant interlayer pairing. Solid and dashed lines denote spectra near a point impurity and the bare spectra (without impurity), respectively. Panels (a) and (b) show the LDOS for pure interlayer pairing, with the impurity located on the outer layer and inner layer, respectively. Panels (c) and (d) correspond to the same impurity positions as in (a) and (b), but with an additional, smaller intralayer pairing component included. }
	\label{fig2}
\end{figure}

We now investigate the single impurity effect for different pairing states and impurity positions. We first focus on the interlayer pairing dominant case. The corresponding LDOS spectra, both in the absence of an impurity and in the vicinity of an impurity, are shown in Fig.~\ref{fig2}. Here, the parameter 
$l$ from Eq. (11) specifies the layer in which the impurity is located.
Figs.~\ref{fig2}(a) and~\ref{fig2}(b) present the results for pure interlayer pairing with $(V_\parallel, V_\perp) = (0, 0.8)$, where the impurity is located on the outer layer ($l=1$) and inner layer ($l=2$), respectively. Figs.~\ref{fig2}(c) and~\ref{fig2}(d) display the spectra for the coexistence case with $(V_\parallel, V_\perp) = (0.6, 0.8)$.

Without the impurity, the bare LDOS exhibits a two-gap structure, which arises from multiband effects. The outer layer shows a larger low-energy residual LDOS, with the zero-energy LDOS being approximately $0.5$. In contrast, the inner layer displays a much smaller residual LDOS, with the zero-energy value being nearly zero. This behavior can be understood by analyzing the layer contribution to the normal-state Fermi surface~\cite{ref10,supp}. Specifically, there exist normal-state Fermi pockets that are entirely contributed by the outer layers. For interlayer pairing, these pockets cannot be paired, resulting in unpaired quasiparticles in the outer layer. As a result, the bare LDOS in the inner layer decreases more sharply near zero energy.

In the presence of an impurity, our results demonstrate that the LDOS spectra near the impurity depend strongly on the impurity position. When the impurity is situated on the outer layer, as shown in Figs.~\ref{fig2}(a) and~\ref{fig2}(c), the LDOS is enhanced inside the superconducting gap, indicating the presence of in-gap states; however, no impurity-induced resonance states are observed. In contrast, when the impurity is located on the inner layer, as seen in Figs.~\ref{fig2}(b) and~\ref{fig2}(d), pronounced resonance peaks appear near zero energy and at a certain negative energy. Notably, the inclusion of a smaller intralayer pairing component does not qualitatively change these results.

\begin{figure}
	\centering
	\includegraphics[width=\linewidth]{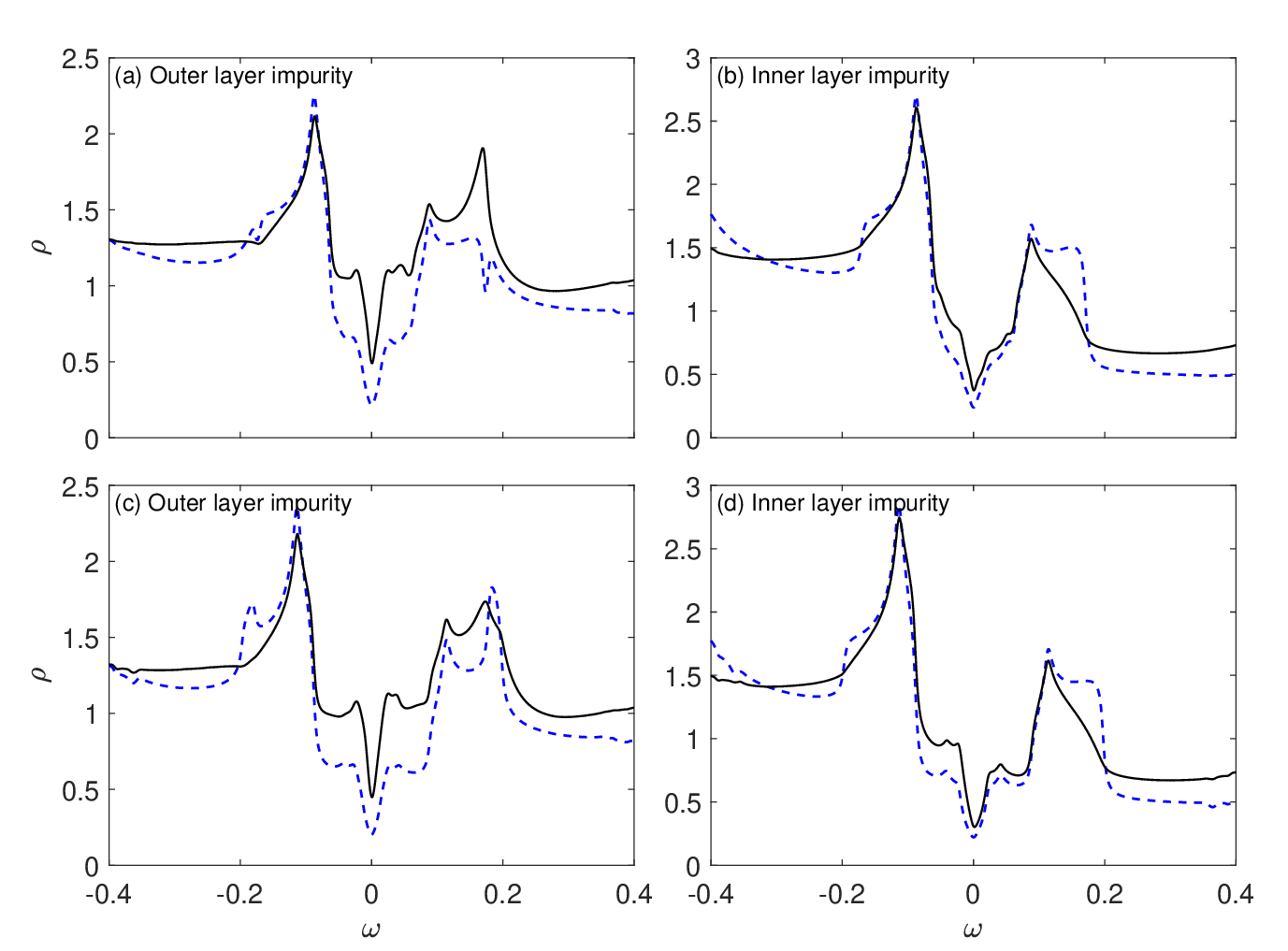}
	\caption{Similar to Fig.~\ref{fig2}, but illustrating the scenario in which intralayer pairing dominates. }
	\label{fig3}
\end{figure}

We now turn to the study of the single impurity effect in the case where intralayer pairing is dominant. The LDOS spectra without an impurity and near an impurity are presented in Fig.~\ref{fig3}. Specifically, Figs.~\ref{fig3}(a) and \ref{fig3}(b) correspond to the pure intralayer pairing case [$(V_\parallel, V_\perp) = (0.8, 0)$], while Figs.~\ref{fig3}(c) and \ref{fig3}(d) show the coexistence case with $(V_\parallel, V_\perp) = (0.8, 0.4)$. 

Residual low-energy LDOS exists due to the presence of quasiparticle nodes in the superconducting state. In the presence of an impurity, as shown in Fig.~\ref{fig3}, the intensities of the LDOS spectra at low energies inside the superconducting gap are enhanced for all intralayer pairing dominant cases, indicating the existence of in-gap states. The intensities of these in-gap states are stronger for the outer layer than for the inner layer. However, no sharp resonance peaks are observed, in contrast to the interlayer pairing dominant states.

\begin{figure}
	\centering
	\includegraphics[width=\linewidth]{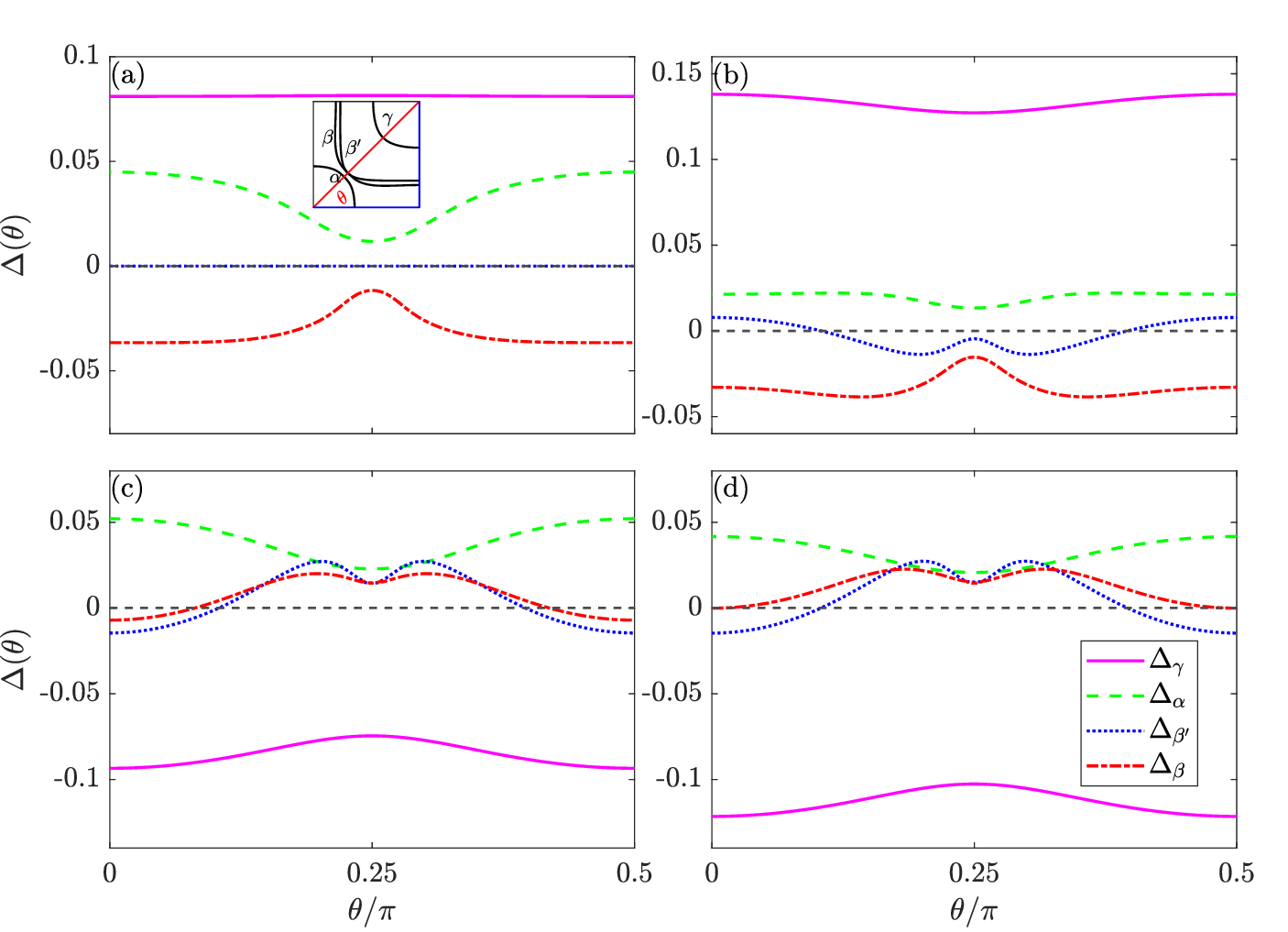}
	\caption{Superconducting gap distribution along the normal-state Fermi surfaces for various pairing interaction parameters $(V_\parallel,V_\perp)$: (a) $(0,0.8)$; (b) $(0.6,0.8)$; (c) $(0.8,0)$; and (d) $(0.8,0.4)$.
	}
	\label{fig4}
\end{figure}

The existence of low-energy in-gap states is, in principle, due to the sign reversal of the superconducting gap along the normal state Fermi surface~\cite{PhysRevB.111.174525}. The superconducting gaps along the normal state Fermi surface for different parameter sets are displayed in Fig.~\ref{fig4}. When the interlayer pairing is dominant [Figs.~\ref{fig4}(a) and \ref{fig4}(b)], the order parameter is positive along the $\gamma$ and $\alpha$ Fermi pockets and negative along the $\beta$ Fermi pocket. For pure interlayer pairing [Fig.~\ref{fig4}(a)], the gap is zero along the $\beta^\prime$ pocket, and the maximal gap magnitudes appear at the $\gamma$ Fermi pocket. In the presence of a smaller intralayer pairing component [Fig.~\ref{fig4}(b)], the gap along the $\beta^\prime$ pocket is nonzero but remains small. The gap magnitude along the $\gamma$ pocket increases. This is because the intralayer order parameters are negative, as shown in Fig.~\ref{fig1}.

For pure interlayer pairing, the entire $\beta^\prime$ pocket remains gapless. This can be understood from the layer-resolved normal-state electronic structure~\cite{supp}: the $\beta^\prime$ pocket originates solely from quasiparticles in the outer layers~\cite{supp}, which cannot form Cooper pairs in an interlayer pairing state. The resulting near-zero gap produces a substantial low-energy bare LDOS in the outer layers. Consequently, these layers exhibit higher residual low-energy LDOS and suppressed superconducting coherence peaks at low energies, as seen in Figs.~\ref{fig2}(a) and \ref{fig2}(c).

When the intralayer pairing is dominant, as seen in Figs.~\ref{fig4}(c) and \ref{fig4}(d), the superconducting gap is positive along the $\alpha$ pocket and negative along the $\gamma$ pocket. For the $\beta$ and $\beta^\prime$ pockets, multiple quasiparticle nodes exist, and the superconducting gaps along the $\beta$ pocket are tiny near the Brillouin zone boundaries. This structure gives rise to significant low-energy residual states in the bare LDOS, again suppressing the low-energy coherence peaks, as illustrated in Fig.~\ref{fig3}.

\begin{figure}
	\centering
	\includegraphics[width=\linewidth]{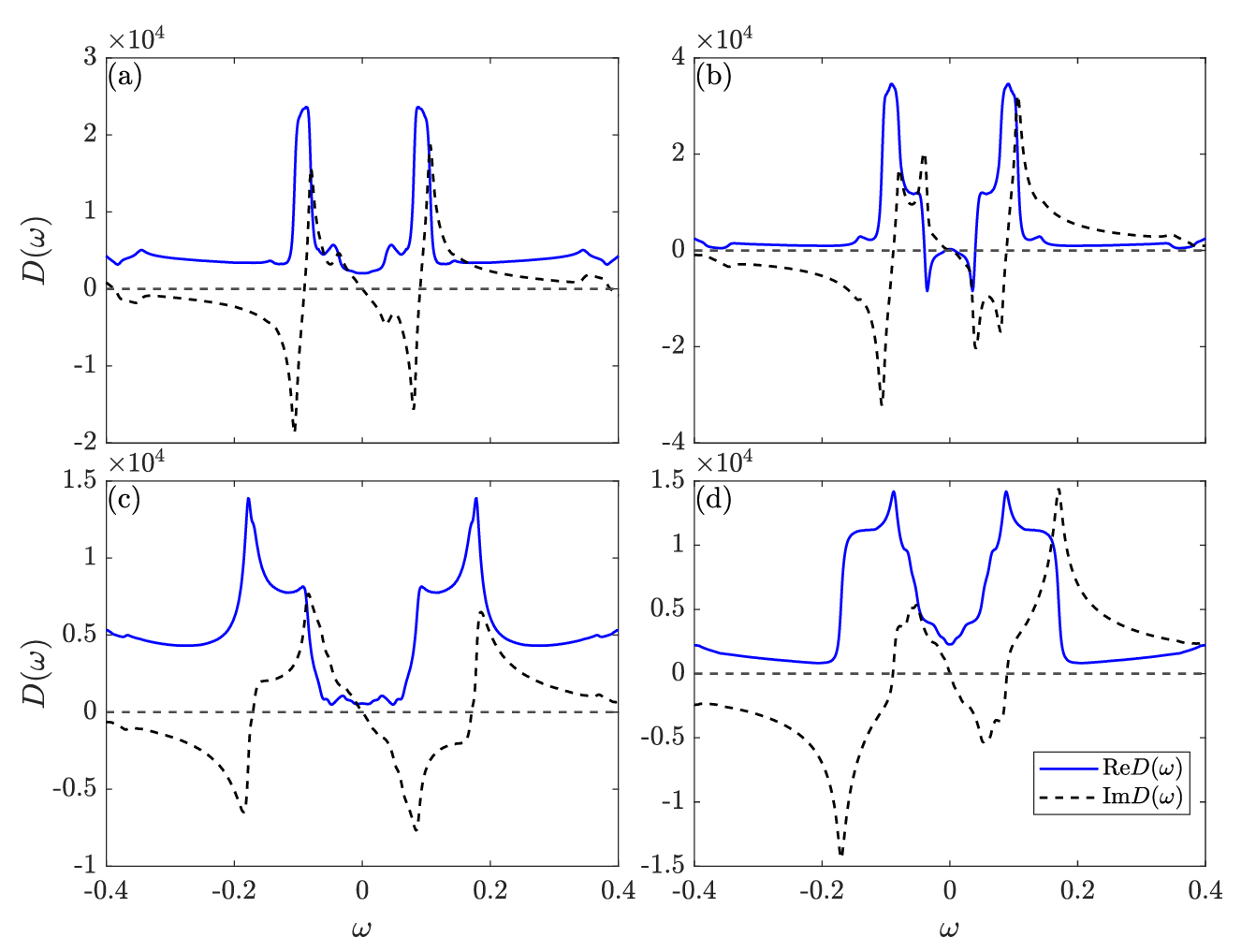}
	\caption{Real and imaginary parts of the function $D(\omega)$
		for pure interlayer pairing (upper panels) and pure intralayer pairing (lower panels). The left and right panels correspond to the outer-layer and inner-layer impurities, respectively.
	}
	\label{fig5}
\end{figure}

With the $T$-matrix method, the impurity states are determined by the denominator of the $T$-matrix, given by
\begin{equation}
    D(\omega) = \det[\hat{I} - \hat{U} \hat{G}_0(\omega)].
\end{equation}
$D(\omega)$ is a complex function, and its real and imaginary parts for pure interlayer pairing and pure intralayer pairing are shown in Fig.~\ref{fig5}.

As can be seen, for all cases, the imaginary part of $D(\omega)$ [Im $D(\omega)$] is an odd function of $\omega$ and vanishes at $\omega=0$. Due to the sign change of the order parameter, 
the real part is generally small at low energies, resulting in the presence of in-gap states~\cite{PhysRevB.111.174525}. 
Specifically, for interlayer pairing, when the impurity is located on the inner layer, the function $\mathrm{Re}\,D(\omega)$ crosses zero near $\omega=0$, leading to a sharp resonant peak close to zero energy, as observed in Fig.~\ref{fig2}~\cite{PhysRevB.100.205119,supp}. However, when the impurity is located on the outer layer or when the intralayer pairing is dominant, at low energies, the pole condition $\mathrm{Re}\,D(\omega)=0$ is not satisfied. As a result, the resonant peak does not exist~\cite{supp}.

While this work examines point impurity effects in the trilayer nickelate superconductor La$_4$Ni$_3$O$_{10}$, several critical challenges remain for future research. First, the current model is based on a non-interacting Hamiltonian. A deeper understanding of impurity effects in this material could be gained by incorporating electronic correlations and investigating how they influence the formation and characteristics of in-gap states. Second, exploring alternative impurity models, such as Anderson-type impurities, could provide crucial new insights into how specific impurity properties affect the superconducting state. Addressing these questions will be essential for advancing our understanding of impurity effects in nickelate superconductors.

\section{SUMMARY}

In summary, we investigate the impurity-induced states in the trilayer high-temperature superconductor $\mathrm{La}_4 \mathrm{Ni}_3 \mathrm{O}_{10}$, a member of the Ruddlesden-Popper nickelate family recently discovered to exhibit superconductivity under pressure. We employ a two-orbital model and $T$-matrix formalism to analyze how single impurities influence the local density of states (LDOS) under different superconducting pairing mechanisms, with particular focus on the distinction between intralayer and interlayer pairing channels.

Due to the asymmetric trilayer structure of $\mathrm{La}_4 \mathrm{Ni}_3 \mathrm{O}_{10}$, quasiparticles near the Fermi level are unevenly distributed between the inner and outer $\mathrm{NiO}_2$ layers, resulting in layer-dependent LDOS. In the interlayer pairing-dominant regime, unpaired quasiparticles remain in the outer layers, leading to a substantial residual LDOS at low energies, while the inner layer remains nearly fully gapped. Our study shows that impurity effects depend strongly on both the pairing symmetry and the impurity position: interlayer-dominant pairing produces sharp resonance peaks near zero energy when the impurity resides in the inner layer, whereas only a general in-gap LDOS enhancement occurs for impurities in the outer layer. In contrast, for intralayer-dominant pairing, in-gap states appear in both layers without the formation of sharp resonance peaks.

These results demonstrate that single-impurity spectroscopy can serve as an effective tool to probe and distinguish the underlying superconducting pairing mechanisms in trilayer nickelate superconductors. They underscore the crucial role of the trilayer structure and multiorbital effects in shaping the superconducting properties of $\mathrm{La}_4 \mathrm{Ni}_3 \mathrm{O}_{10}$, providing valuable guidance for future experimental investigations into the nature of superconductivity in layered nickelates.


%

\renewcommand{\thesection}{S-\arabic{section}}
\setcounter{section}{0}  
\renewcommand{\theequation}{S\arabic{equation}}
\setcounter{equation}{0}  
\renewcommand{\thefigure}{S\arabic{figure}}
\setcounter{figure}{0}  
\renewcommand{\thetable}{S\Roman{table}}
\setcounter{table}{0}  
\onecolumngrid \flushbottom 
\newpage

\begin{center}\large \textbf{Supplemental material for layer-resolved impurity states reveal competing pairing mechanisms in trilayer nickelate Superconductor La$_4$Ni$_3$O$_{10}$} \end{center}

\section{Impurity effect with different impurity strengths}

\renewcommand \thefigure {S\arabic{figure}}
\begin{figure}[htb]
\centering
\includegraphics[width = 18cm]{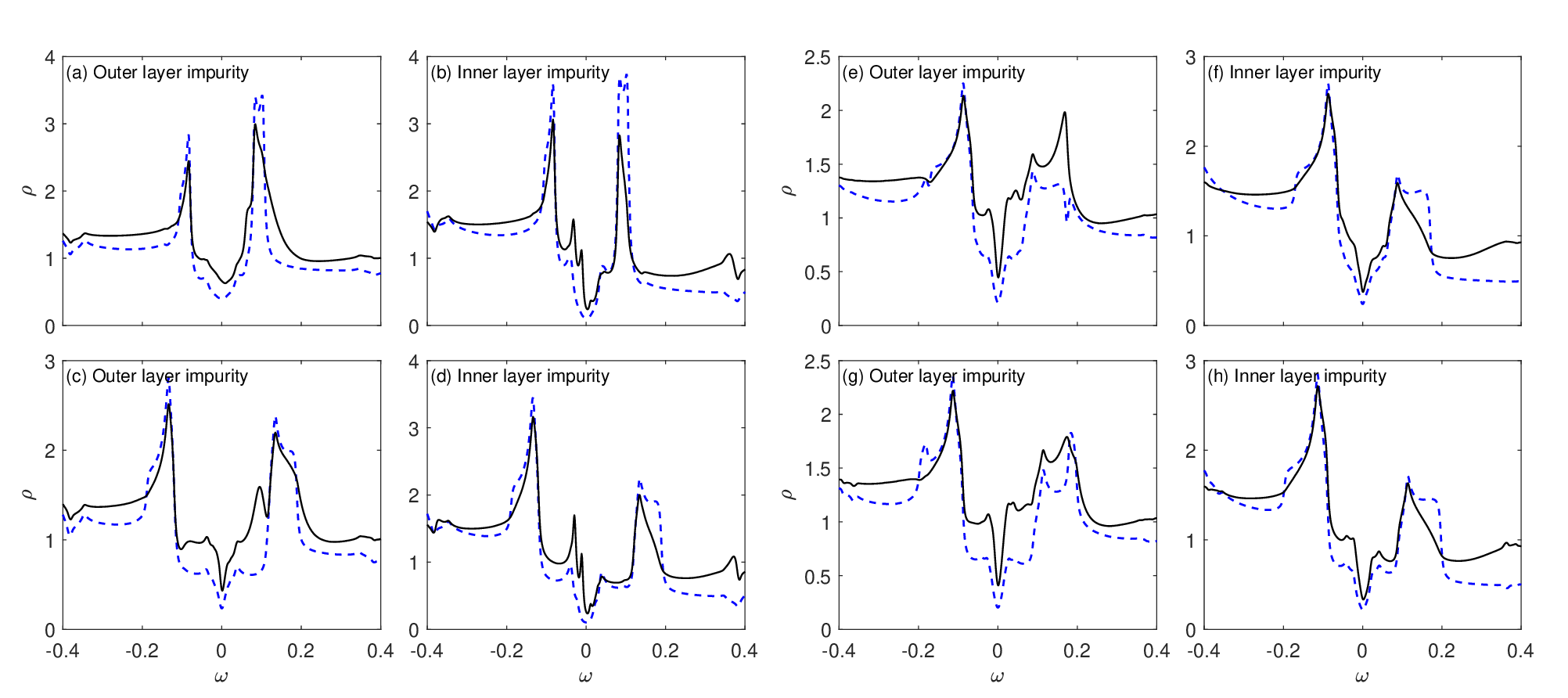}
\caption{\label{figs1} LDOS spectra for an impurity strength $V_i=3$. The configurations in (a–d) and (e–h) follow Fig. 2 (dominant interlayer pairing) and Fig. 3 (dominant intralayer pairing) of the main text, respectively.}
\end{figure}

\begin{figure}[htb]
\centering
\includegraphics[width = 18cm]{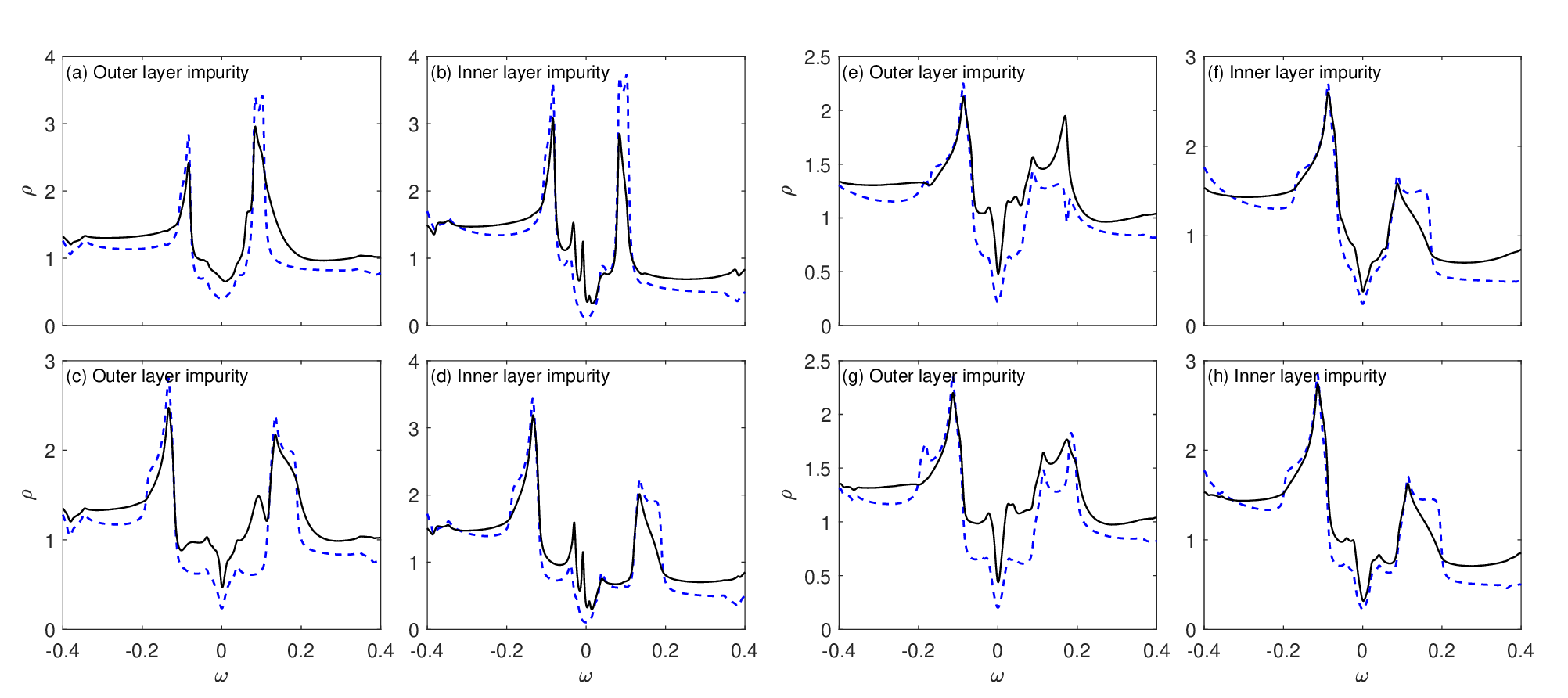}
\caption{\label{figs2} Similar to Fig. S1, but with the impurity strength $V_i=5$.}
\end{figure}

\begin{figure}[htb]
\centering
\includegraphics[width = 18cm]{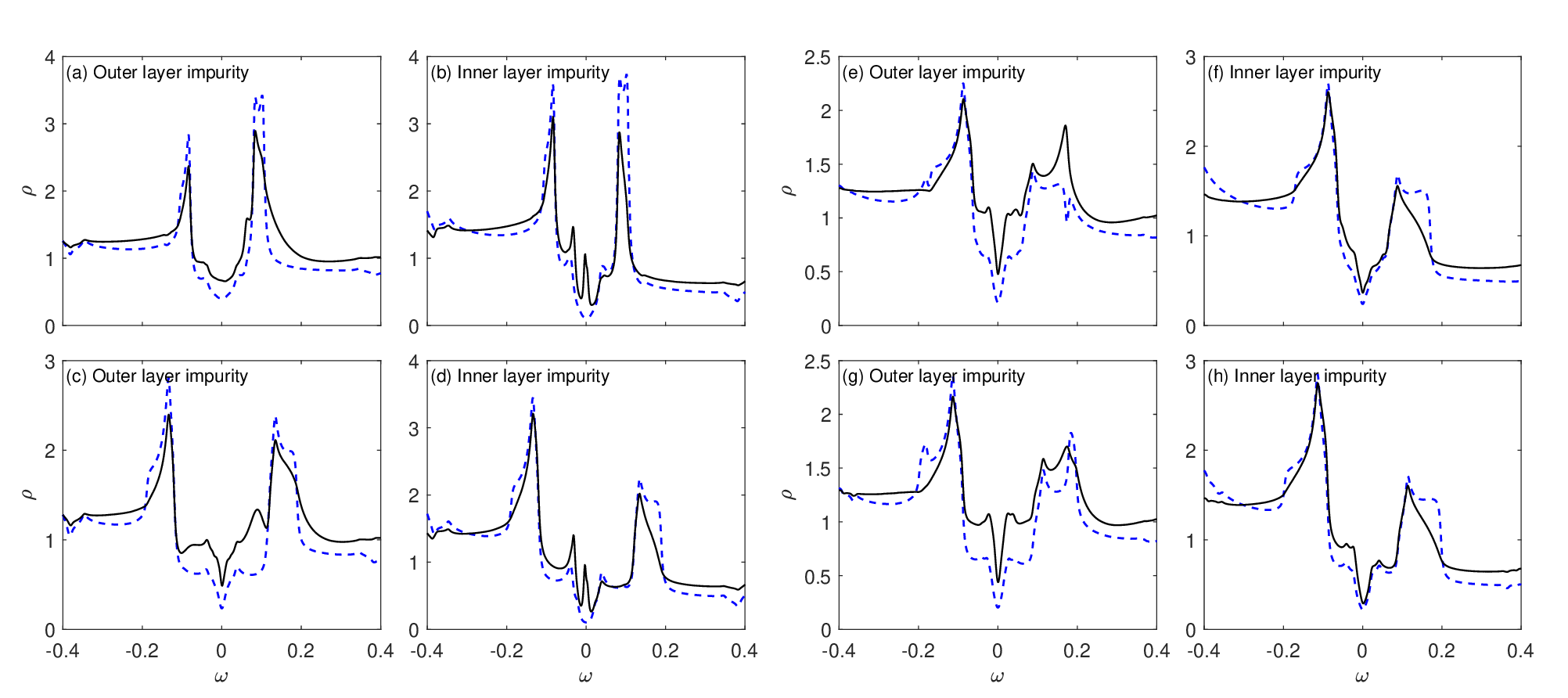}
\caption{\label{figs3} Similar to Fig. S1, but with the impurity strength $V_i=50$.}
\end{figure}

The main text discusses LDOS spectra near an impurity with strength 
$V_i=10$, revealing sharp in-gap resonant states under conditions of dominant interlayer pairing and an inner-layer impurity. To assess the robustness of this phenomenon, we extended our numerical analysis to a wider range of impurity strengths (Figs. S1–S3). These results demonstrate qualitative consistency across different impurity potentials. Notably, the low-energy in-gap states survive for all impurity strengths investigated whenever interlayer pairing is dominant and the impurity is situated on the inner layer. This confirms that the reported effect is a robust feature of the system.

\section{Layer and orbital resolved zero-energy normal state spectral function}

The superconducting gap structure for pure interlayer pairing, shown in Fig.~4(a) of the main text, can be understood by examining the layer- and orbital-dependent spectral function at zero energy. The normal-state Hamiltonian is expressed as a $6\times 6$ matrix, $H = \sum_{\mathbf{k}} \hat{\Psi}_{\mathbf{k}}^{\dagger} \hat{H}_{\mathbf{k}} \hat{\Psi}_{\mathbf{k}}$, with
\begin{align}
\hat{\Psi}_{\mathbf{k}}=(c_{\mathbf{k} x \uparrow}^{1}, c_{\mathbf{k} x \uparrow}^{2}, c_{\mathbf{k} x \uparrow}^{3}, c_{\mathbf{k} z \uparrow}^{1}, c_{\mathbf{k} z \uparrow}^{2}, c_{\mathbf{k} z \uparrow}^{3} )^{T}.
\end{align}

The spectral function for layer $l$ is obtained by diagonalizing the Hamiltonian matrix $\hat{H}_{\mathbf{k}}$:
\begin{equation}
A_l(\mathbf{k}, \omega) = -\frac{1}{\pi}\sum_{n} \frac{|u_{ln}({\bf k})|^2+|u_{l+3,n}({\bf k})|^2}{\omega - E_{n}(\mathbf{k}) + i\Gamma},
\end{equation}
where $E_{n} (\mathbf{k})$ are the eigenvalues of the Hamiltonian and $u_{l n} (\mathbf{k})$ are the corresponding eigenvector components

Similarly, the orbital-dependent spectral function is given by
\begin{equation}
A_\tau(\mathbf{k}, \omega) = -\frac{1}{\pi}\sum_{n}\sum^3_{l=1} \frac{|u_{\tau+l,n}({\bf k})|^2}{\omega - E_{n}(\mathbf{k}) + i\Gamma},
\end{equation}
where $\tau=0$ and $3$ correspond to the $d_{x^2-y^2}$ and $d_{z^2}$ orbitals, respectively.

\begin{figure}[htb]
\centering
\includegraphics[width = 12cm]{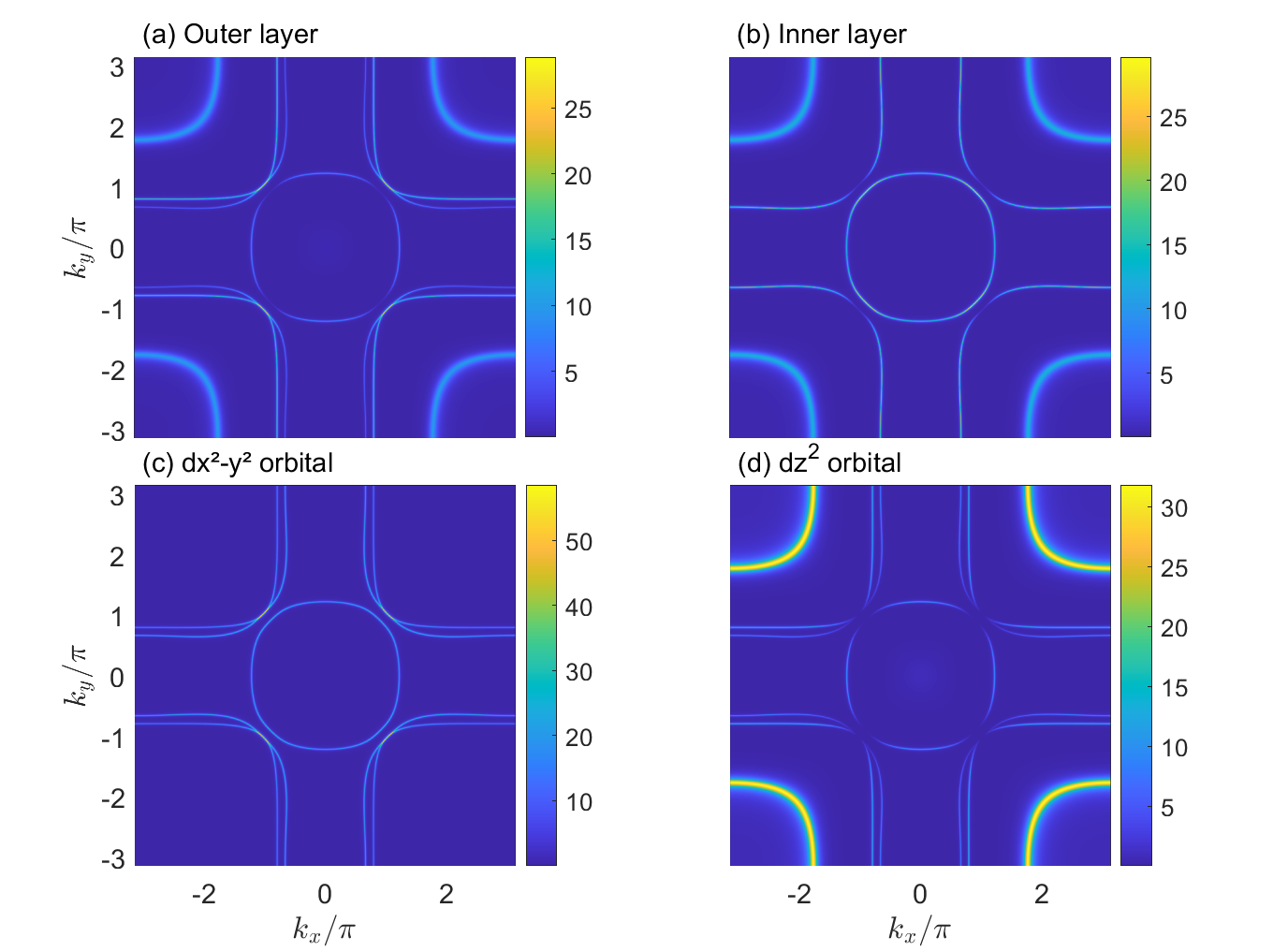}
\caption{\label{figs4} Intensity plots of layer and orbital resolved zero energy normal state spectral function.}
\end{figure}

Fig.~S4 presents intensity plots of the layer- and orbital-dependent zero-energy spectral function. Since the spectral function is maximal at the Fermi momentum, these plots reveal the layer and orbital weights of the normal-state Fermi surface. As shown in Figs.~S4(a) and S4(b), the $\beta^\prime$ Fermi pocket derives entirely from the outer layer. Consequently, this pocket remains ungapped for pure interlayer pairing. The other three Fermi pockets ($\alpha$, $\beta$, $\gamma$) have contributions from both the outer and inner layers, allowing interlayer pairing to generate nodeless energy gaps around them. The sign structure of these gaps is consistent with that in the bilayer material La$_3$Ni$_2$O$_7$, as understood from a simplified bilayer model~\cite{PhysRevB.111.174525}.

The anisotropy of the energy gap around the $\alpha$ and $\beta$ pockets can be understood from their orbital composition. Figs.~S4(c) and S4(d) show that along the diagonal direction, these pockets are dominated by the $d_{x^2-y^2}$ orbital. Moving away from the diagonal, the weight of the $d_{z^2}$ orbital increases. Given that the order parameter magnitude is larger for the $d_{z^2}$ orbital than for the $d_{x^2-y^2}$ orbital (see Fig.~1 in the main text), the gap magnitude consequently increases away from the diagonal. In contrast, the $\gamma$ pocket is composed solely of the $d_{z^2}$ orbital, resulting in an almost isotropic energy gap.

\section{Further analysis of the $T$-matrix denominator}

In the main text, we analyzed the denominator of the $T$-matrix, defined as $D(\omega) = \det[\hat{I} - \hat{U} \hat{G}_0(\omega)]$, a complex function whose examination clarifies the formation of impurity-induced resonant states.

The imaginary part of $D(\omega)$ can be expressed as
\begin{equation}
\mathrm{Im}D(\omega) = -\det[U\mathrm{Im} G_0(\omega)].
\end{equation}
For an impurity located in the outer layer, the matrix $U$ has non-zero elements $U_{11} = U_{44} = -U_{77} = -U_{10,10} = V_i$. Due to particle-hole and time-reversal symmetry, $\mathrm{Im}D(\omega)$ simplifies to $-V_i\mathrm{Im}[G_0(\omega) - G_0(-\omega)]$, where $G_0(\omega)$ is the bare single-particle Green's function for the relevant layer. Consequently, $\mathrm{Im}D(\omega)$ is generally an odd function of $\omega$.

The real and imaginary parts of $D(\omega)$ are related via the Kramers-Kronig transformation:
\begin{equation}
\mathrm{Re}D(\omega)=\frac{1}{\pi} P \int d \omega^\prime \frac{\mathrm{Im}D(\omega^\prime)}{\omega^\prime-\omega},
\end{equation}
where $P$ denotes the principal value. For a function obeying this relation, a smooth imaginary part implies a smooth real part. However, an abrupt change in $\mathrm{Im}D(\omega)$ at a specific frequency induces a logarithmic divergence in $\mathrm{Re}D(\omega)$. To illustrate, consider a step-function form $\mathrm{Im}D(\omega)=A\theta (\omega-\omega_0)$. The Kramers-Kronig relation then yields
\begin{equation}
\mathrm{Re}D(\omega)=\frac{1}{\pi} P \int^{\infty}_{\omega_0} d \omega^\prime \frac{A}{\omega^\prime-\omega}= -\frac{A}{\pi}\ln |\omega_0-\omega|.
\end{equation}

Applying this insight to the physical system: when intralayer pairing is dominant, the superconducting gaps on the $\beta$ and $\beta^\prime$ pockets are small. This leads to residual low-energy LDOS and suppressed coherence peaks in the superconducting state. Consequently, $\mathrm{Im}D(\omega)$ varies smoothly at low energies without abrupt changes, resulting in a smooth $\mathrm{Re}D(\omega)$ that cannot satisfy the resonance condition $\mathrm{Re}D(\omega)=0$.

In contrast, when interlayer pairing dominates, the low-energy LDOS is significantly different between layers. As noted in the main text, the outer layer retains a large number of unpaired quasiparticles, which suppresses its low energy coherence peaks. The inner layer, however, displays well‑defined coherence peaks at low energies. Consequently, for an impurity placed in the inner layer, the imaginary part of the response function, $\mathrm{Im}D(\omega)$ drops sharply at a specific low energy. This drop induces a corresponding sharp dip (logarithmic feature) in $\mathrm{Re}D(\omega)$. Since $\mathrm{Re}D(\omega)$ is typically positive at zero energy~\cite{PhysRevB.111.174525}, this dip forces it to cross zero at a nearby energy, creating the condition for an impurity-induced in-gap resonant peak.

\end{CJK}
\end{document}